\documentclass[useAMS,usenatbib]{mn2e}
\usepackage{epsfig}
\title[]
{Planet Migration through a Self-Gravitating Planetesimal Disk}
\author[]
{Alex Moore,  Alice C. Quillen, \& Richard G. Edgar
\\
Department of Physics and Astronomy, University of Rochester,
Rochester, NY 14627, USA   \\
}

\begin{document}
\label{firstpage}
\maketitle

\begin{abstract}
We simulate planet migration caused by interactions between
planets and a planetesimal disk.  
We use an N-body integrator optimized for near-Keplerian motion 
that runs in parallel on a video graphics card, and that
computes all pair-wise gravitational interactions.
We find that the fraction of planetesimals found in mean motion resonances 
is reduced and planetary migration rates are on average about 50\% slower
when gravitational interactions between the planetesimals are computed
than when planetesimal self-gravity is neglected.  
This is likely due to gravitational stirring
of the planetesimal disk that is not present 
when self-gravity is neglected that reduces their capture
efficiency because of the increased particle eccentricity dispersion.
We find that
migration is more stochastic when the disk is self-gravitating
or comprised of more massive bodies.
Previous studies have found that if the planetesimal disk density
is below a critical level, migration is ``damped'' and the
migration rate decays exponentially, otherwise
it is ``forced'' and the planet's migration rate could accelerate exponentially.
Migration rates measured from our undamped simulations suggest that 
the migration rate saturates at a level proportional to disk density
and subsequently is approximately power law
in form with time.

\end{abstract}

\section{Introduction}

The field of solar system dynamics, has 
a timeline of discoveries that is related to the computational power available
(e.g., \citealt{morby01}).
As computational power has increased over time, so to has our ability 
to more accurately simulate more complex systems. 
The interaction between planets and planetesimals subsequent to formation
is a well posed n-body problem but displays remarkable complexity  
even in the absence of collisions,
including resonance capture, planetary migration, and heating and scattering of planetesimals.
In this paper we attempt to more accurately simulate this system
by using the increased computational power and on board memory recently
available on video graphics cards.

Planet-planetesimal scattering can lead to planet migration 
(e.g., \citealt{fernandez84,malhotra95,hahn99,ida00,levison03,gomes04,gomes05,hahn05,
levison08}).  
Recent work on dusty circumstellar disks with clearings, suggest that planets are 
required to account for the morphology of the disk \citep{quillen06,wyatt06}.  The 
planetesimal masses suggested by collisional models 
(e.g., \citealt{wyatt02,DD03,quillen07}) and proximity of the hypothetical planet 
to the disks for systems such as Fomalhaut intimate that planetary migration could 
be taking place.  To simulate planet migration due to planetesimal scattering, 
planets must gravitationally interact with planetesimals. However, because of the 
$O(N^2)$ required computational intensity previous simulations have necessarily neglected 
the gravitational interactions between the planetesimals 
(e.g., \citealt{hahn99,gomes04,hahn05,thommes08}). This means that collective gravitational
effects and gravitational self stirring in the disk have been ignored.  Here we take all 
interparticle forces into account to explore what differences might be seen in 
simulations of migrating planets with and without planetesimal interactions.  

\section{Numerical simulations}

The migration simulations were carried out using an N-body code running on a 4 
node cluster running the "Mars Hill" Rocks 4.3 Linux operating system (a 
CentOS based distribution) that hosts 4 NVIDIA GeForce 8800 GTX graphics cards.  
The code was run on the graphics processing units (GPUs) residing on the graphics
cards and is written with NVIDIA's CUDA (Compute Unified Device Architecture), 
a C-language development environment for CUDA enabled GPUs. CUDA is a GPGPU 
(General-purpose computing on graphics processing units) technology that allows 
a programmer to use the C programming language to code algorithms for execution 
on the graphics processing unit.  Its intention is to expose the hardware to 
the developer through a memory management model that encourages both constant
streaming of data as well as massive parallelization.  

\subsection{A second order democratic heliocentric method symplectic integrator
for the GPU}

We have modified the second order symplectic integrator introduced by 
\citep{duncan98}, also known as the democratic heliocentric method, so that it runs
on a GPU.  We have chosen the democratic heliocentric method because the force from 
the central body is separated from the integration of the remainder of the particles
and the coordinates do not depend on the order of the particles.  This increases the 
accuracy of the integrator when there are large mass differences and is particularly 
desirable when forced to work in floating point single precision.  When we obtain 
video cards that can compute in double precision the code will improve in precision.  
The move to double precision will also allow us to extend the range of the 
planetesimal masses simulated. 

In heliocentric coordinates and barycentric momenta \citep{wisdom96} 
the Hamiltonian of the system can be written 
\begin{equation}
H = H_{Sun} + H_{Kep} + H_{Int}
\end{equation}
where 
\begin{equation}
H_{Sun} = {1 \over 2 m_0} \left|\sum_{i=1}^n  {\bf P}_i^2\right|^2 
\end{equation}
is a linear drift term and $P_i$ are the barycentric momenta.
Here $m_0$ is the central particle mass.
The second term
$H_{Kep}$ is the sum of Keplerian Hamiltonians for all particles with
respect to the central body,  
\begin{equation}
H_{Kep} = \sum_{i=0}^N \left ({ {{\bf P}_i^2 \over 2 m_i} 
- {G m_i m_0  \over \left| {\bf Q}_i \right|}} \right)
\end{equation}
where ${\bf Q}_i$ are the heliocentric coordinates and are conjugate
to the barycentric momenta.  Here $m_i$ is the mass of the $i$-th particle
and $G$ is the gravitational constant.
The interaction term 
contains all gravitational interaction terms except those to the central body,
\begin{equation}
H_{Int} = \sum_{i=1}^N \sum_{j=1,j\ne i}^N -{ G m_i m_j \over 2 \left|{{\bf Q}_i - {\bf Q}_j}\right|}.
\end{equation}

The second order single timestep integrator advances with timestep $\tau$
using evolution operators (e.g., \citealt{yoshida90})
\begin{equation}
E_{Sun} \left({\tau \over 2}\right)
E_{Kep} \left({\tau \over 2}\right)
E_{Int} \left({\tau  }\right)
E_{Kep} \left({\tau \over 2}\right)
E_{Sun} \left({\tau \over 2}\right)
\end{equation}
where we have reversed the order of the Keplerian evolution
and the interaction steps compared to that discussed by \citep{duncan98}.
We have done this to reduce the total number of computations per timestep. 
The Keplerian advance requires $O(N)$ computations but interaction term 
requires $O(N^2)$ computations.

The drift evolution operator requires computation of the sum of the momenta.  
We have implemented this using a parallel reduction sum algorithm available 
with the NVIDIA CUDA Software Development Kit (SDK) 1.1 that is similar to 
the scan prefix sum algorithm \citep{harris08}.  

The Keplerian step was 
implemented with $f$ and $g$ functions using the universal differential 
Kepler's equation \citep{prussing93} so that bound and unbound particles can
both be integrated with the same routine.  The Keplerian evolution step is 
also done on the GPU with each thread computing the evolution for a separate 
particle.  
The dominant source of error is in the Keplerian evolution step
and is due to the single floating point precision.
These errors can
cause a systematic radial drift that does not average to zero.
To minimize errors caused by the single
precision computation during the Kepler advances we chose
$f$ and $g$ functions that maintain angular momentum conservation across each  
evolution step.
The positions and velocities at a later
time can be written in terms of those at an earlier time
\begin{eqnarray}
\vec{{\bf x}}_1 &=& f \vec{{\bf x}}_0 + g \vec{{\bf v}}_0 \nonumber \\
\vec{{\bf v}}_1 &=& \dot{f} \vec{{\bf x}}_0 + \dot{g}  \vec{{\bf v}}_0.
\end{eqnarray}
The angular momentum at the later time, $\vec{\bf L_1}$ can be written
in terms of that at the earlier time, $\vec{\bf L_0}$,
\begin{equation}
\vec{{\bf L}}_1 = (f \dot{g} - g \dot{f})(\vec{{\bf x}}_0 \times \vec{{\bf v}}_0)
                = (f \dot{g} - g \dot{f}) \vec{{\bf L}}_0.
\end{equation}
Conservation of angular momentum 
yields the condition
\begin{equation}
f \dot{g} - g \dot{f} = 1.
\end{equation}
We utilize this formula to solve for
one of the 4 functions reducing
the inward radial drift resulting
by the single precision computation during the Keplerian advances.  

The interaction terms are computed on the GPU with all $N^2$ force pairs 
evaluated explicitly in parallel.  The algorithm is based on the algorithm 
described by \citep{nyland08}.  This algorithm takes advantage of fast 
shared memory on board the GPU to simultaneously compute all forces
in a $p \times p$ tile of particle positions, where $p$ is the number of 
threads chosen for the computation (typically 256).  The total energy was 
evaluated with a kernel explicitly evaluating all $N^2$ pair potential 
energy terms, similar to that calculating all $N^2$ forces.

After the change to heliocentric/barycentric coordinates, the position of 
the first coordinate corresponds to the center of mass and center of momentum.  
The trajectory of this particle need not be integrated.  However it is convenient
to calculate the energy using all pair interactions including the central mass.  
The interaction term in the Keplerian part of the Hamiltonian can be computed 
at the same times as $H_{Int}$ if ${\bf Q}_0$ is set to zero.  Consequently we set 
${ \bf Q}_0 = {\bf P}_0 = 0$ at the beginning of the computation.  This is
equivalent to working in the center of mass and momentum reference frame.
Because we would like to be able to quickly check the total energy, we have 
chosen to keep the first particle corresponding to the center of mass and 
momentum as the first element in the position and velocity arrays.  During 
computation of $H_{Int}$ we set $m_0$ to zero so that force terms from the 
first particle are not computed. These are already taken into account in the
evolution term corresponding to $H_{Kep}$.  The mass is restored during the 
energy sum computation as all potential energy terms must be calculated explicitly.

The location of memory should be considered when running routines as 
data transfer between the CPU and GPU can slow the code.  The maximum 
theoretical throughput of PCI-express 16x bus technology, the interlink
between the CPU and GPU on Intel processor based motherboards, is 4 GB/s 
with a significantly higher latency than on device memory transfers.  Depending on 
GPU design, theoretical memory transfer throughputs can approach 100 GB/s, as 
is the case for the 8800 GTX which has a theoretical throughput of 86.4 GB/s.  
Reducing the number of data transfers between the CPU and GPU not only 
reduces the CPU clock time required for a calculation, but also reduces 
the number of calls which have higher latency.  CPU and GPU global memory 
access is a significant bottleneck compared to the actual number of clock 
cycles required for a specific calculation with the former containing a much 
higher performance penalty due to latency.  For this reason, positions and 
velocities for all particles are kept in global memory on board the GPU device.  
Arrays for both current and previous time step positions are required during the 
interaction step computation.  The particle positions and velocities are 
only transferred back into host or CPU accessible memory to output data 
files.  An additional float vector of length equal to the number of particles
is allocated in global memory on the device to compute the momentum sums 
used in the drift step computation.  Shared memory on the GPU is used 
during the all pairs interaction computation step and during the reduction 
sum.  Shared memory on the GPU drastically increases the speed of calculations 
because the use of shared memory has almost no latency penalty compared to 
that of both the CPU and the GPU global memory.  By streaming information 
from global memory to shared memory, we are able to hide the latency of global 
memory, further increasing the computation speed.  Though the maximum number 
of threads on the video card we used is 512, we found that register space on 
the GPU limited the interaction step computation to 256 threads per block due 
to its complexity.  A more detailed review of NVIDIA GPU hardware and 
programming techniques can be found in the CUDA Programming Guide.

We work in a 
lengthscale in units of the outermost planet's initial semi-major axis and with 
a timescale such that $GM_* = 1$ where $M_*$ is the mass of the central star.

\subsection{Code Checks}

%As with any new piece of software, we spent a large amount of time debugging 
%as much of the code as possible.  
The following self-consistency checks on the 
code were performed. We checked that two body dynamics is preserved for all 
particles when all masses but the central one are zero. In this case the 
orbital elements (excepting the mean anomaly) are conserved at the level of 
the precision of computation.  We checked that the energy computation computed
in heliocentric/barycentric coordinates is consistent with that computed in 
Cartesian coordinates.  We checked that the conversion to heliocentric/barycentric 
coordinates followed by inverse conversion returns the initial coordinates.  
We checked that the force and potential energy terms computed on the GPU are 
consistent with those computed on CPU.  We checked that the N-body simulation 
without Keplerian integration conserves energy.  We note that the order
of some of the arguments in the SDK1.1 nbody kernel were inconsistent with the 
correct order described by \citep{nyland08}.  To conserve energy we corrected
the order of the arguments in the force computation step from that
in the SDK1.1 nbody kernel.
We checked that integrating a two 
body system conserves energy.  We ran a test case identical to that run by 
\citet{bromley06} and shown in their Figure 5 with similar results (though see 
discussion on energy conservation in the next section).  This corresponds to 2 
earth mass planets embedded in a uniform disk of 800 objects 1/100 of the mass 
of the planets.  This test case was also integrated by \citet{kokubo95}.

\subsection{Simulation Checks}

The second set of checks was done by running full simulations in order to determine 
the smoothing length and time step requirements and restrictions.  In one test we 
varied the time step significantly in order to measure the impact of larger time 
steps on our accuracy.   A larger timestep would allow us to run the simulation
for more orbital periods. 
We ran full $10^4$ particle simulations 
over thousands of orbits and noted that varying the time step resulted in little difference
in the total energy measured.  
In our second test, we varied the softening length in 2 different sets of simulations.  
At extremely small softening length values we note that the simulated system
becomes extremely stochastic.  The reverse effect is exhibited when we make the 
smoothing length extremely large.  This effect was previously noted in \citep{kokubo95} 
and happens because our simulation does not adjust the time step to take into account 
collisions and close encounters.  Since the simulation steps are 
small enough to allow "smooth" or nearly continuous updates, there is the 
possibility that a close encounter between two objects will be forced by the step 
which may have never occurred physically.  If the softening length is made too large, 
the forces between objects that are having close encounters is reduced unrealistically.  
The softening length is chosen to help prevent this from happening. 
%Guidelines on  how to select an appropriate softening length 
%are discussed by \citet{kokubo95}.  
\citet{kokubo95} selects a softening parameter set as 
\begin{equation}
R_{min} = { r_{H} \over 200 }  
\end{equation}
where $r_{H}$ is the Hill radius of a planetesimal. 
This leads to a Hill radius which is roughly equivalent to the physical radius of 
a protoplanet at 1AU.  Given these parameters, we chose a variety of smoothing lengths
that were on the same order of magnitude and did not note a significant any 
significant deviations between the results of the simulations.  The smoothing 
length and timestep were not adjusted during the simulation.

\subsection{Benchmarks}

We measured the time to run our code on on the NVIDIA GeForce 8800 GTX graphics cards. 
The timing included the delay to send the job out to the individual nodes of the cluster, 
initial disk generation and input and output of files.  We ran a simulation of a disk 
with 2 planets, one at semi-major axis $a=1$ with the other interior to this.  The 
planet masses were $10^{-3}$ of that of the star.  The planets were surrounded by a 
disk of particles with mass $10^{-7}$ of that of the central star.  Time units were 
set such that $GM_*=1$.  The time step was 0.1 or $0.1/(2\pi)=0.0159$ of
the outer planet's orbit.
For $10^4$ particles 1000 timesteps ran with 0.137 s per timestep.
For $10^5$ particles 100 timesteps ran with 57s per timestep and similar error.
The energy fraction error was $10^{-3}$ throughout the entire simulation and drifts 
downward systematically with time such that the system becomes increasingly bound.
The systematic loss of energy and associated radial inward drift we have determined 
is caused by the single point floating precision for calculations done during the 
Kepler advances.

\subsection{Initial conditions}

In this paper we primarily discuss simulations with 2 planets and an exterior
annulus of planetesimals.
The initial planetesimal disk has a flat distribution in semi-major axis so that the 
surface density is proportional to $1/r$ for radius $r$.  
An isotropic disk is generated with 
initial eccentricity and inclination dispersion such that
$\langle e^2\rangle = 2 \langle i^2 \rangle$ 
and initial inclination dispersion set to 0.01.  
The smoothing length was typically set to be slightly smaller 
than the mean interparticle spacing.  Typical mean spacing values for $10^4$ particles 
in units of the inner planet's semi-major axis was around 0.01, 
with our smoothing length set correspondingly to 0.01.  
The smoothing length spacing was set the same for all simulations.
The timestep was set at 0.1 where the orbital period of 
the inner planet is $2 \pi$.  
A timestep of 0.1 is convenient because it ensures 
accuracy while still allowing of order $10^3$ orbital periods to be completed within 
24 hours for $10^4$ particles.
Comparisons between simulations with
timesteps of 0.2, 0.1 and 0.05 yielded 
only insignificant differences in planetary migration rates, 
indicating that our simulations are not
strongly sensitive to the chosen time step. All other 
parameters are listed in table \ref{tab:tab1}.  The initial semi-major axes of the 
planetesimals ranges from 1.5 to 3.0.
We work in units of the outermost planet's initial semi-major axis and with 
$GM_* = 1$ so that the initial period of the outermost planet is $2\pi$.  We discuss 
most timescales in units of the outermost planet's initial rotation period.  
The initial inclination and eccentricity of the two planets were set to zero.
However, we have found  that planet eccentricities and inclinations 
rapidly settled toward zero for the planetesimal disks considered here.
All simulations were run for approximately 4000 orbits.

%10^4 particle simulation with self gravity

\begin{figure*}
\begin{center}
\includegraphics[angle=0,width=6.8in]{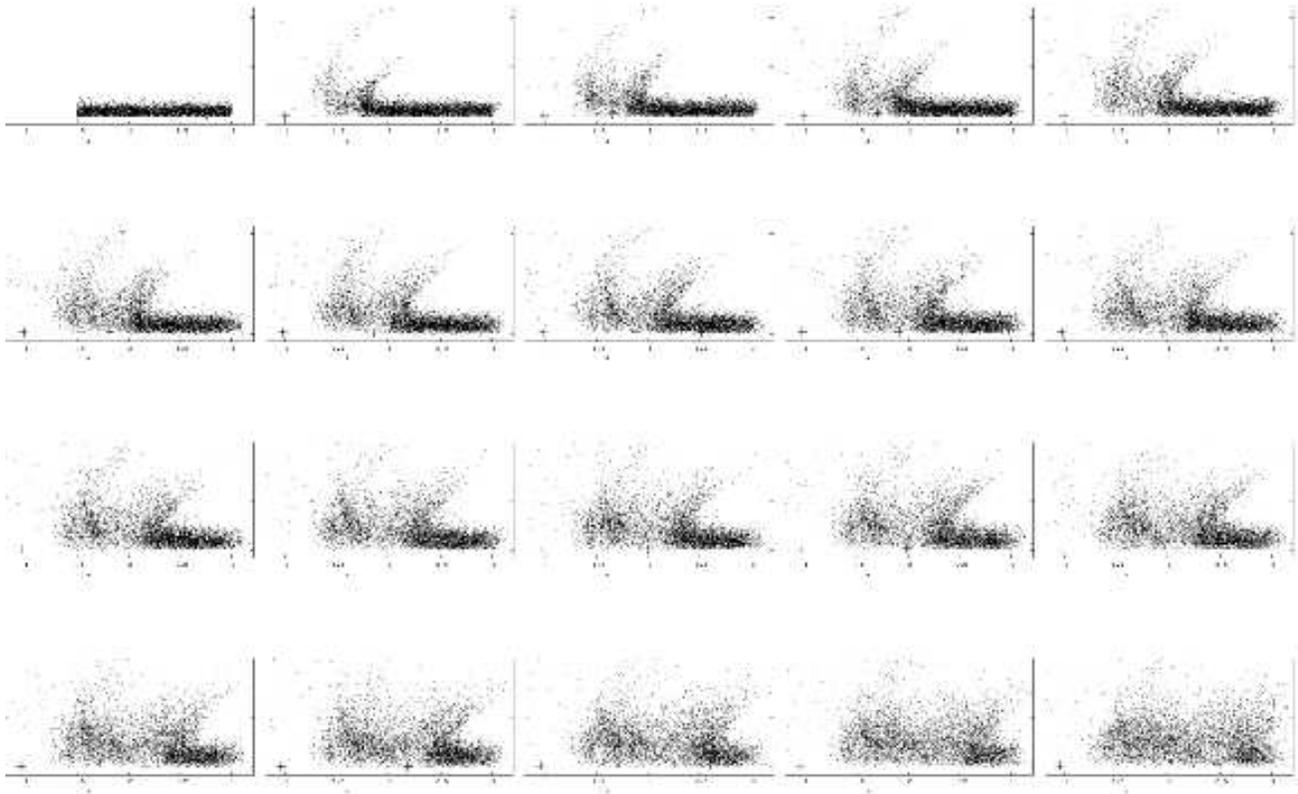}
%\vspace{-0.25cm}
\caption
{The semi-major axis ($x$-axis) vs eccentricity ($y$-axis) distribution of particles
and planets
different times during simulation C1.  Each hash on the y axis is an increment 
of 0.2 in eccentricity starting at 0.  Each hash on the x-axis is an increment 
of 0.5 in semi major axis, with the inner planet starting at 1.0 and the planetesimals
in the disk taking an initial starting position between 1.5 and 3.0.  Our migrating 
planet has an initial semi major axis of 1.5.  This simulation has
10000 particles and computes all force pairs so that the disk
feels self-gravity.  Each frame is separated by 200 rotation periods 
based on the migrator's initial orbital period. 
The two planets are shown as green crosses and the planetesimals
as red dots. Two scattering surfaces are seen, one associated
with each planet.  These are broad features with eccentricity
and semi-major axis given a particle a planet orbit crossing periastron. 
In the first half of the simulation the eccentricities of particles
in the outer disk increases due to self-gravity.  Eccentricities
increase for particles that are not scattered outward as the planet
passes.
\label{fig:10000}
}
\end{center}
\end{figure*}

%10^4 particle simulation with no self gravity

\begin{figure*}
\begin{center}
\includegraphics[angle=0,width=6.8in]{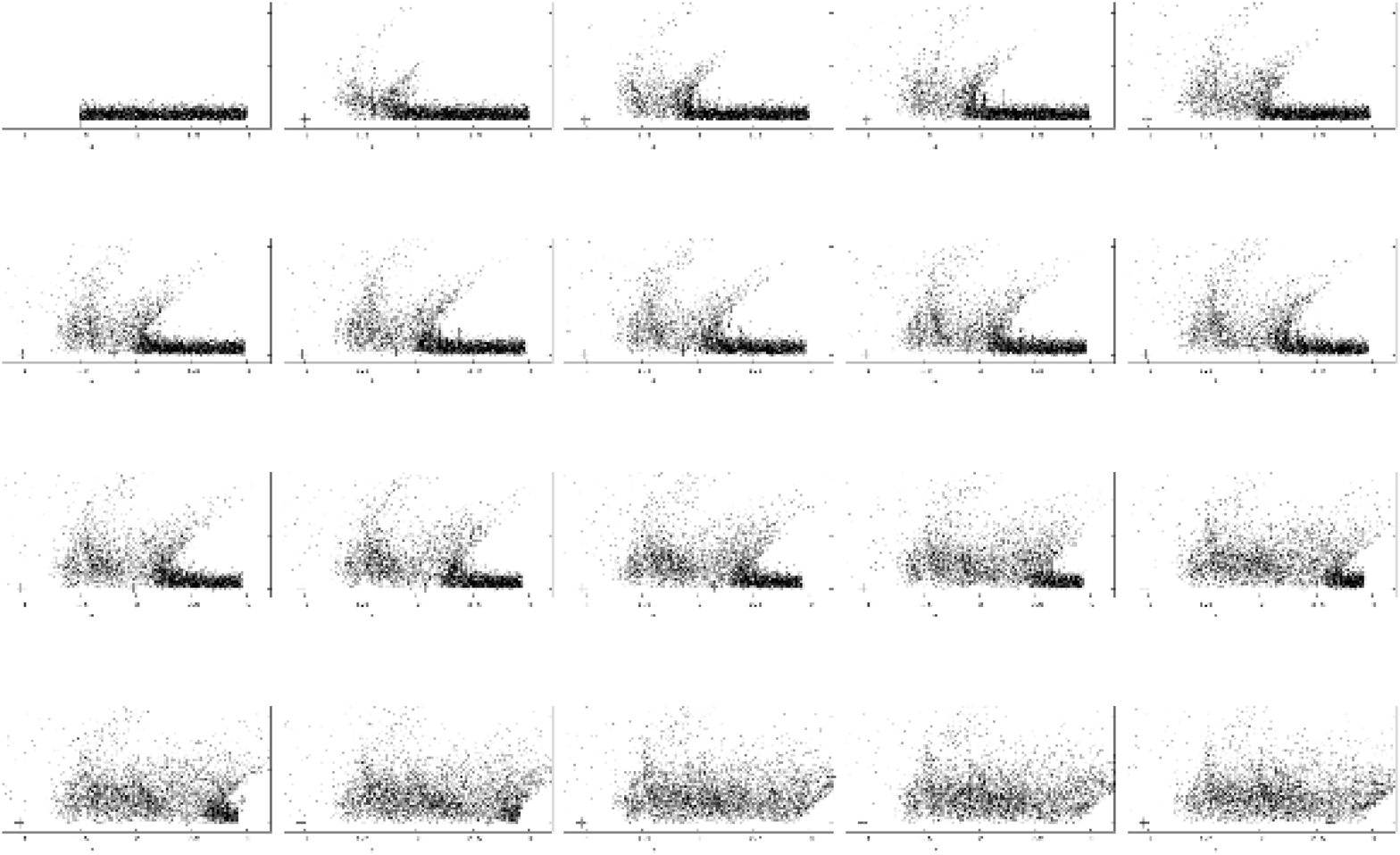}
%\vspace{-0.25cm}
\caption
{Similar to Figure \ref{fig:10000} except showing the simulation N1 
that is identical to C1 but lacking
self-gravity in the disk.  Each hash on the y axis is an increment 
of 0.2 in eccentricity starting at 0.  Each hash on the x-axis is an increment 
of 0.5 in semi major axis, with the inner planet starting at 1.0 and the planetesimals
in the disk taking an initial starting position between 1.5 and 3.0.  Our migrating 
planet has an initial semi major axis of 1.5.  Similar scattering surfaces
are seen in this simulation. The outer planet migrates more quickly
outwards than in simulation C1.   The outer disk remains colder 
(has a lower eccentricity dispersion) during
the first half of the simulation compared to
the simulation with self-gravity in the disk. 
Populations of particles trapped in mean motion resonances with
the planet are more prevalent in this simulation than
in simulation C1.
\label{fig:10000-noselfgrav}
}
\end{center}
\end{figure*}

\subsection{Comparison code lacking forces between planetesimals}

Our code can compute all interparticle forces, however
most planet migration simulations have neglected forces between planetesimals.
We desire a direct comparison between simulations that take into
account all inter particles forces and those that neglect forces
between planetesimals.   To do this we run two versions of our
code that are identical except that interparticle forces are not
computed if both particle masses are below  that of a planet.
In all other respects the two sets of simulations are identical.

\section{Migration of 2 planet systems}

We begin our study of migration with the simplest system that allows migration of the 
outer planet into the disk; that of two planets just inside an initially cold 
(low velocity dispersion) annulus of planetesimals.  
Our first set of simulation has a inner planet with mass
$10^{-3}$ of that of the central star (similar to Jupiter) and an outer planet
with a mass ratio of $5 \times 10^{-5}$ (similar to Uranus). 
A low mass was initially chosen for the outer planet 
so that outer disk is not completely disrupted during migration.
A mass similar to that of Uranus was chosen simply because in 
our test runs it showed the most interesting variation in
its migration rate.  Larger planet masses showed increasing disk 
disruption while lower masses showed greater degrees of 
stochastic motion.
A high mass inner planet was initially chosen to ensure that particles
can be efficiently scattered by the outer planet and subsequently ejected
by the inner planet.  While this mechanism can function for an inner
planet that has mass ratio less than $10^{-3}$, the migration of
the outer planet is then more
sensitive to the distance between the planets.  Our choice of
a large inner planet reduces the dependence on the distance between
the planets. 
If the distance between the two planets is large then it takes
longer for
the outer planet to scatter particles sufficiently that
they cross the inner planet's orbit. 

The parameters of our simulations are listed in table \ref{tab:tab1}.
Three sets of simulations are run.  The first ten, denoted C1-C10, include
all pair-wise forces and contain a range of planetesimal disk masses and number
of planetesimals.  
In these simulations the planetesimal mass is fixed
at $10^{-7}$ of the stellar mass.   
The most massive disks simulated (simulations C1, N1, and V1-V10) have
a mass ratio of $10^{-3}$.   This is a factor of a few
larger than the more massive 200 Earth mass disks simulated
by previous studies \citep{hahn99,gomes04}.
Simulations
N1-N10 are identical to C1-C10 except forces between planetesimals are
neglected.   Simulations V1-V10 compute all force pairs,
have the same total planetesimal disk mass,
but vary the number of planetesimals and the planetesimal mass.

We first discuss the semi-major axis and eccentricity
distributions during the simulations.   We compare simulations with
and without planetesimal self-gravity.
We then discuss differences
seen in the migration rates of the outer planet.
Last, we consider the effects of increasing the number of particles
in a simulation while maintaining a constant mass disk. 

\subsection{Eccentricity and semi-major axis distributions during migration}

Figure \ref{fig:10000} shows the semi-major axis vs eccentricity
distribution  at different times during the C1 simulation 
computing all force pairs for 10000 planetesimals.  Each
frame in the figure is separated by 200 orbital times
where an orbital time corresponds to the initial orbital period
of the outer planet.
As can be seen from the position of the outer planet in
each frame it migrates outward during the simulation.  
The inner planet drifts inwards but not significantly
so because it is more massive than the outer planet.
The particles
with semi-major axis interior to the outer planet 
are higher after the planet has passed than they were
before hand, leaving a possible signature of
a migrating planet (e.g., \citealt{lufkin06}).  
We also notice an increase in the eccentricity
dispersion of particles in the outer disk with time during
the first half of the simulation.

We can compare this simulation to an identical one,  denoted N1 but
lacking planetesimal self-gravity.  The semi-major
axis and eccentricity distributions for this simulation
are shown in Figure \ref{fig:10000-noselfgrav}.
The increase in eccentricity dispersion of particles in
the outer disk seen in the self-gravitating disk
is not present in simulation N1.
Thus the eccentricity dispersion increase in the self-gravitating
outer disk is likely caused 
by gravitational stirring of the planetesimals by themselves.  

Two dominant scattering surfaces are seen as broad strokes
in the semi-major axis vs eccentricity distributions 
of Figure \ref{fig:10000} and \ref{fig:10000-noselfgrav}, 
one each from the inner and outer planet.
These surfaces correspond to particles with semi-major
axis and eccentricity that have
periastron crossing a planet's orbit.
Similar scattering surfaces have previously been seen in simulations of migrating
planets (e.g., \citealt{gomes04,hahn05,thommes08}) 
and imply that both planets are scattering 
planetesimals outwards.  However, for the outer planet to migrate outwards it must 
also scatter planetesimals inwards.  The total energy and angular momentum
of material scattered outwards by the inner planet must exceed that scattered
outward by the outer planet to allow it to migrate outwards.  
This is one of the primary reasons why we 
required a large inner planet mass.
%, and perhaps why we found little 
%migration in  Jupiter and  Saturn planet mass 
%simulations despite the fact that there was more than enough mass in the disk 
%to allow migration of the outer planet.  

When the outer planet migrates outwards it replenishes
a population of moderately eccentric bodies residing between the planets.
This allows the inner planet to continually scatter objects.  As
these objects are scattered outwards, they stop interacting with
the  outer planet allowing it to continue migrating outwards.
Previous studies have described migration
in terms of two regimes (e.g., \citealt{gomes04}). 
If the outer planet migrates at increasing slower rates,
its migration is described as "damped" (e.g., \citealt{ida00}), however if it
continues to migrate until it reaches the edge of the disk
the migration is considered "forced" \citep{gomes04}.
The planetesimal disks considered here are sufficiently massive
that the planet migrates outwards until it is near the edge of
the disk.

A comparison between 
Figure \ref{fig:10000} and \ref{fig:10000-noselfgrav}, 
shows little difference between the morphology of the scattering
surfaces, however the location of
the one associated with the outer planet
is at a larger semi-major axis for simulation N1 than C1.  This
is because the outer planet migrates more quickly in the
simulation lacking disk self-gravity (N1) than that including self-gravity (C1).
We will discuss migration rates in the next subsection.

%\begin{figure}
%\includegraphics[angle=0,width=3.6in]{1000noselfgrav_still.eps}
%\includegraphics[angle=0,width=3.6in]{10000noselfgrav_still.eps}
%\caption{
%\label{fig:1000-10000noselfgrav_still}
%Semi-major axis ($x-$axis) vs eccentricity ($y-$axis) 
%distributions at a particular time during the the N10 10000 particle simulation. 
%A few particles have been captured into the 4:3 and 3:2 resonance at semi
%major axes of 1.82 and 1.96  with
%the outward migrating outer planet at a semi-major of 1.5.
%Particles have been captured into the 4:3 and 3:2 resonances
%at semi-major axes of 2.2 and 2.35 with the outer planet
%with semi-major axis of 1.8.
%}
%\end{figure}

\begin{figure}
%\vspace{-1.5cm}
%\hspace{.85cm}
\includegraphics[angle=0,width=3.5in]{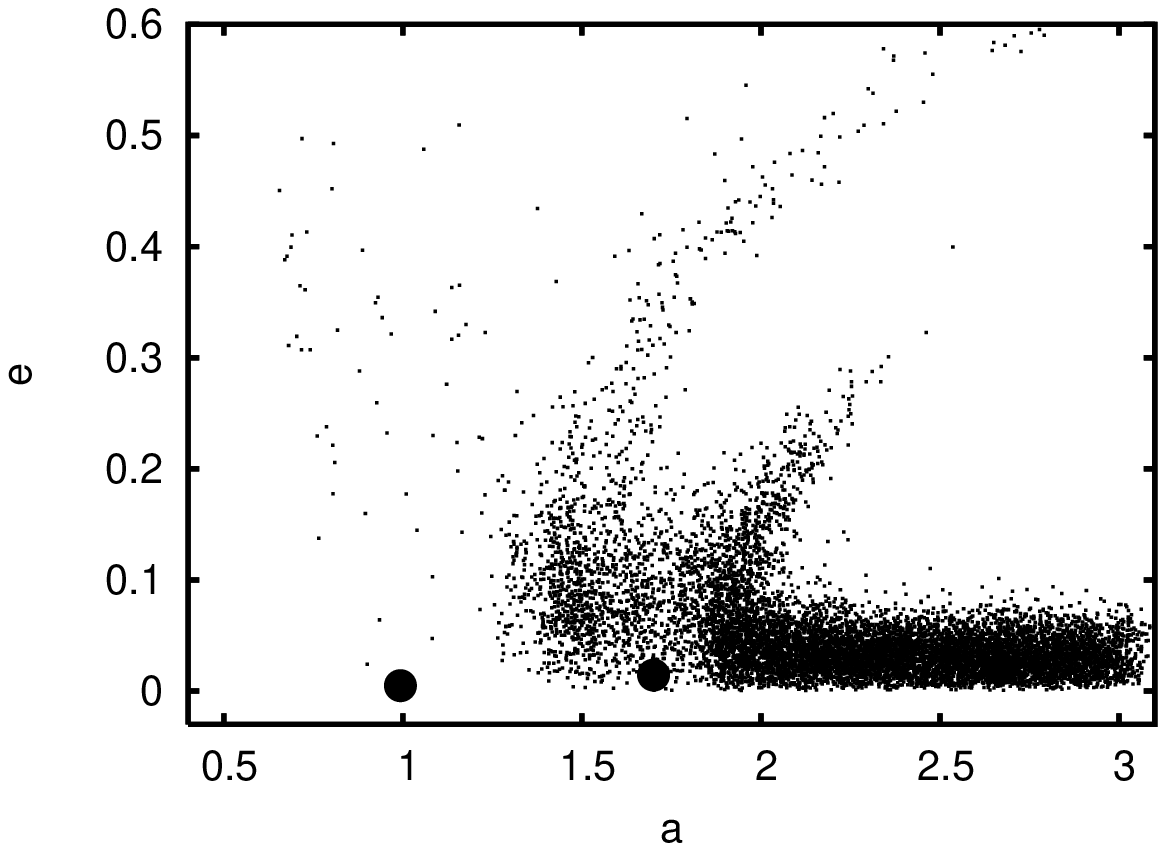}
%\vspace{0.4cm}
%\hspace{.85cm}
\includegraphics[angle=0,width=3.5in]{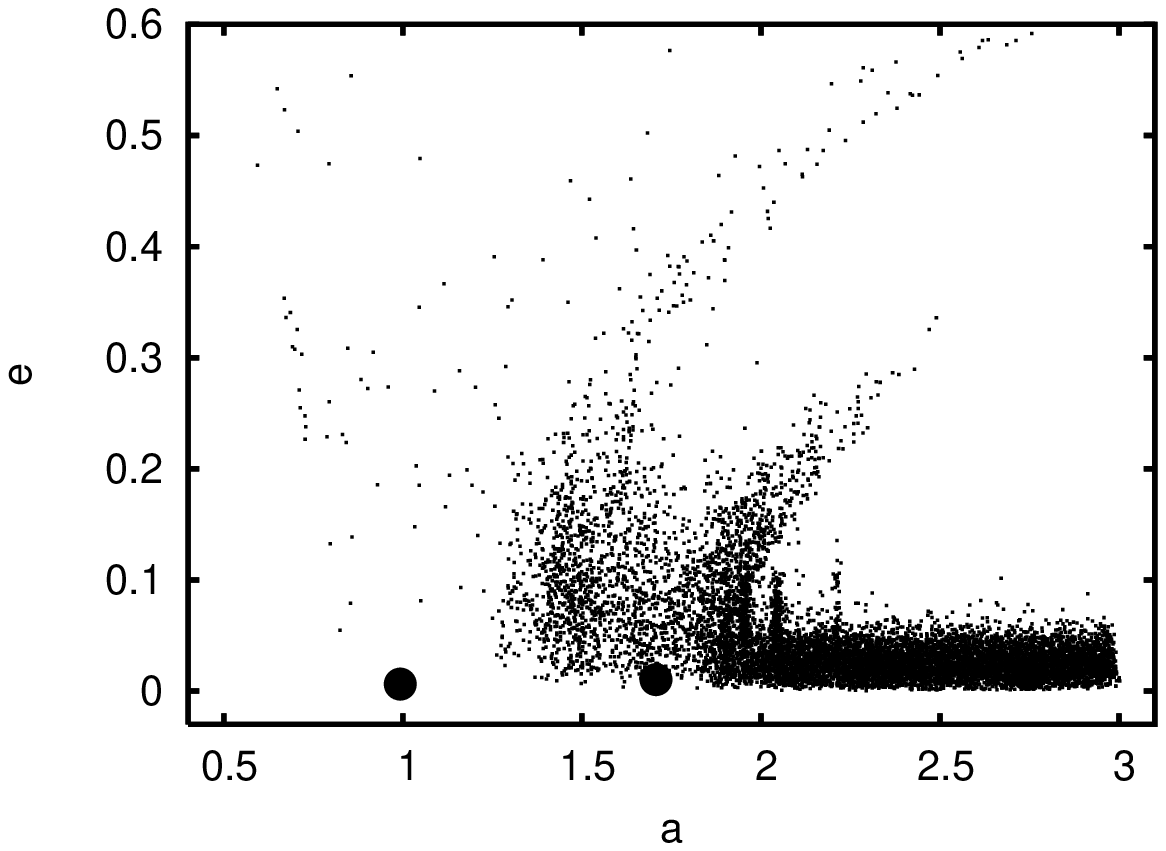}
%\vspace{0.75cm}
\caption{
\label{fig:ae15}
Eccentricity vs semi-major axis distribution
at a time $t=557$ rotation periods.
Planets are shown with filled in circles, planetesimals with small dots.
Scattering surfaces with periastrons that are planet crossing are seen
for both planets.
a) For simulation C1 including planetesimal gravitational interactions.
b) For simulation N1 lacking planetesimal interactions.
In simulation N1 particles have been captured in the 4:3 and 3:2 mean motion 
resonances at semi-major axes of 2.06 and 2.23 for the planet at a semi-major
axis at 1.7.  The outer disk in simulation C1 is thicker and lacks
as numerous a resonant population as simulation N1.
In simulation C1 either planetesimal interactions have scattered
planetesimals out of resonances or the fraction of particles captured
into resonances is reduced because of the higher eccentricity
distribution resulting from gravitational stirring.
}
\end{figure}

A notable difference between the simulations with and
lacking disk self-gravity is the fraction of bodies 
captured into mean motion resonances with the outer planet.  
In figure \ref{fig:ae15} we show simulations C1 and N1 at 557 orbits into
their runs (approximately 15 percent completion).
We find that the fraction of particles captured into resonances is lower
in the simulation with self-gravity than that without.  While an outwardly migrating 
planet is capable of capturing bodies into mean motion resonances, the fraction
of particles captured into resonance can be reduced if the planet
is migrating faster or if the initial eccentricity distribution of
the disk is larger \citep{ida00,hahn05,quillen06b}.  
The lifetime of particles in 
resonances could be reduced by scattering among the planetesimals or
if the planet's motion is stochastic (e.g., \citep{hahn99}). 
Because the planet migration rates up to this
point in the simulation are not significantly different in 
the two simulations, we attribute 
the difference in resonance population to either reduced capture because
of a higher initial eccentricity distribution or
because of a reduced lifetime in resonance.
There are particles present in mean motion resonances
in simulation C1 however the fraction of particles
captured in resonances is small compared to  that in simulation N1 lacking
disk self-gravity.

\subsection{Effects of disk self-gravity and disk and particle mass on the 
planet migration rate}

\begin{figure}
%\vspace{-1.5cm}
%\hspace{.85cm}
\includegraphics[angle=0,width=3.6in]{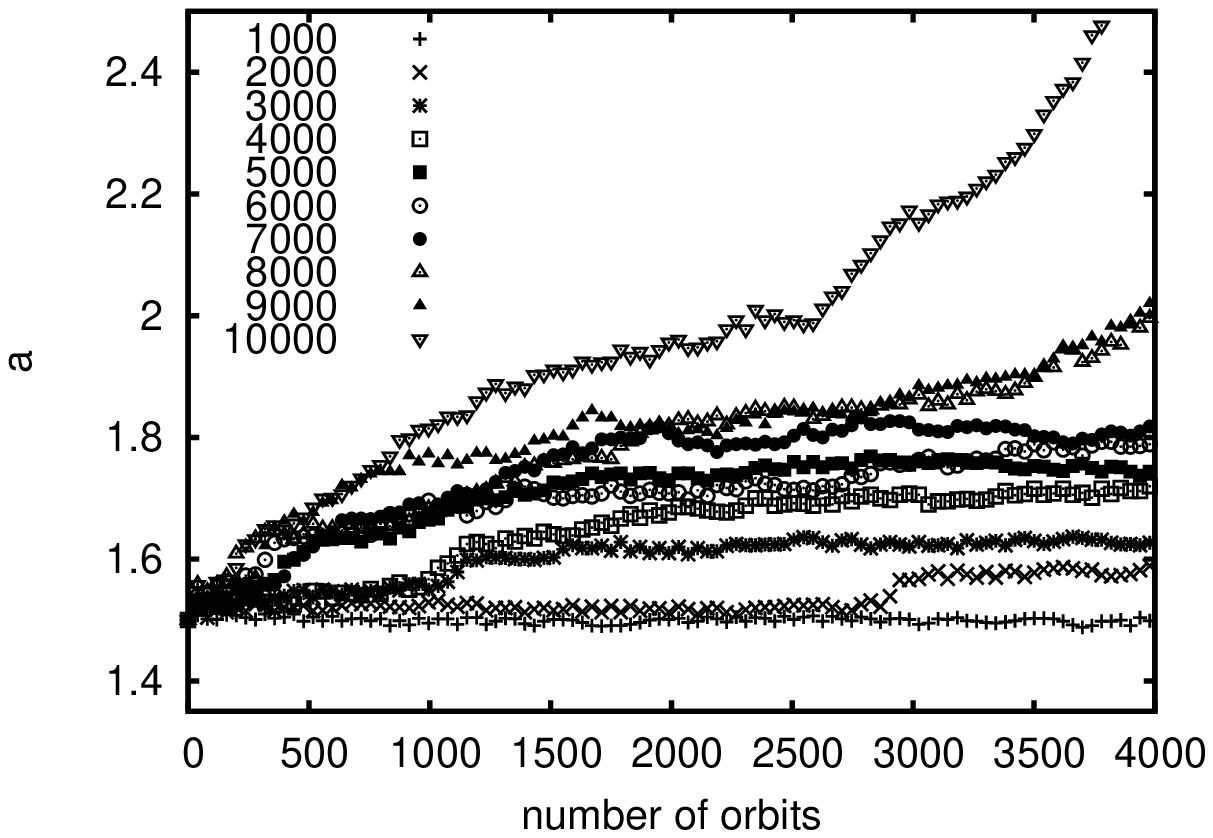}
%\vspace{0.4cm}
%\hspace{.85cm}
\includegraphics[angle=0,width=3.6in]{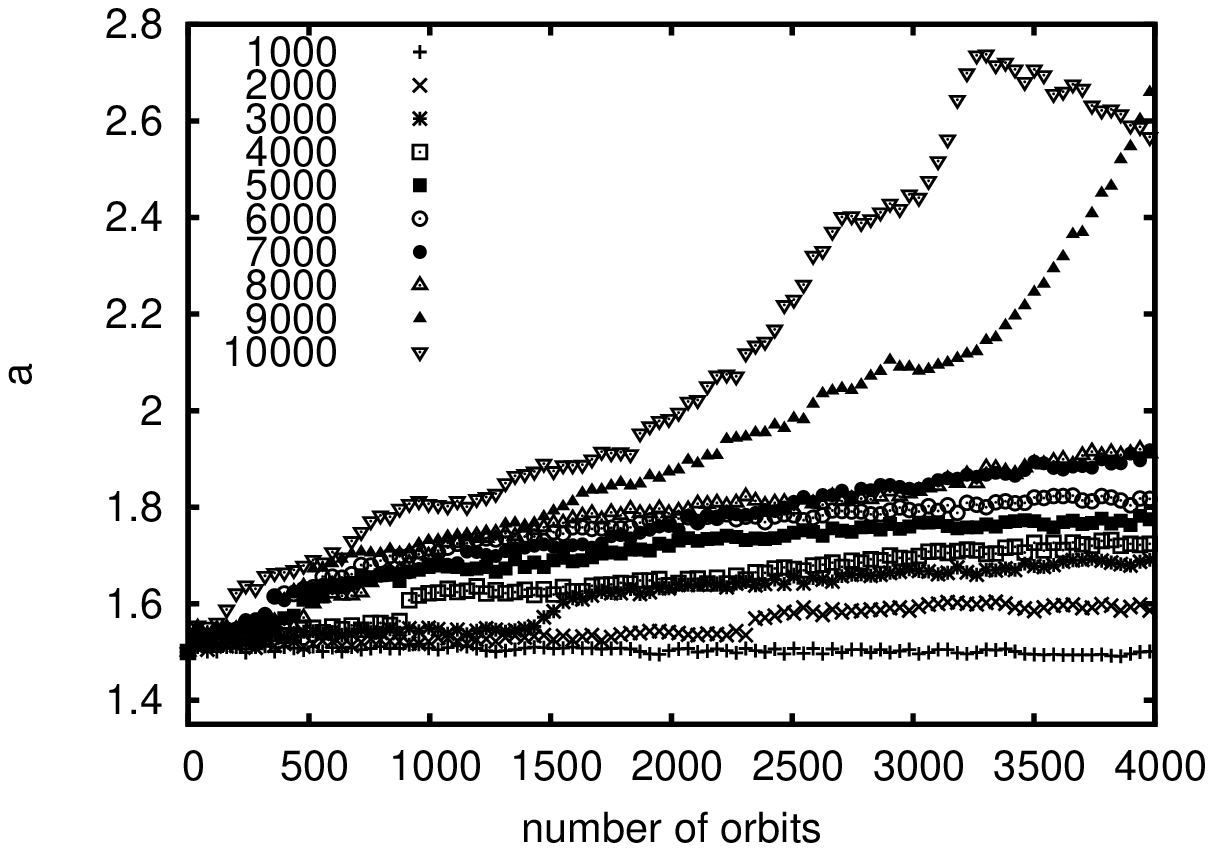}
%\vspace{0.75cm}
\caption{
\label{fig:mig_rate_profile}
The outer planet's semi-major axis as a function of time.
a) For simulations C1 through C10, computing all pair-wise forces.
b) For simulations N1 through C10, lacking disk self gravity.
Jumps in the semi-major axis occur when the two planets pass
through resonance.
The migration rate is approximately proportional to the
disk density.  Stronger migration is seen at the largest
disk densities.   Migration is not damped but neither
is it exponentially increasing in these simulations.
This suggests that the migration rate saturates
at a level that depends on disk density.
}
\end{figure}

%\subsection{Migration rates}

In Figure \ref{fig:mig_rate_profile} we show the semi-major
axis of the outer planet 
as a function of time for simulations C1 - C10 (with disk 
self-gravity) and for simulations N1- N10 (lacking disk self-gravity).
Since the simulations shown in this figure have planetesimals all of the same mass,
the simulations with more particles have higher disk masses.
We find that more massive disks allow the outer planet
to migrate faster, as might be expected from scaling models
predicting migration rates and previous numerical
studies (e.g., \citealt{hahn99,gomes04}).
In Figure \ref{fig:mig_rate_profile} we see that there
are jumps in the semi-major axes profiles with time, particularly
for the lower mass disks with the slowest migration rates.  
These correspond to times when the two planets pass through
a mean motion resonance.  As the planets are separating,
the outer planet is not captured into resonance, rather
it jumps from one side of the resonance
to the other.  For a short time
the planet eccentricities increase and then are damped
via scattering with planetesimals (a form of dynamical friction).  
The jumps in
both eccentricity and semi-major axis, we see in
both the self-gravitating and non-self-gravitating
simulations and are particularly noticeable when
the disk is low mass and the migration rate slow.
These resonant events
have been previously seen in other numerical simulations,
for example such an event is an important characteristic of
the "Nice" model for the outward migration 
of the outer planets in the early solar system
\citep{tsiganis05,levison08}.

\begin{figure}
%\vspace{-1.0cm}
%\hspace{0.85cm}
\includegraphics[angle=0,width=3.6in]{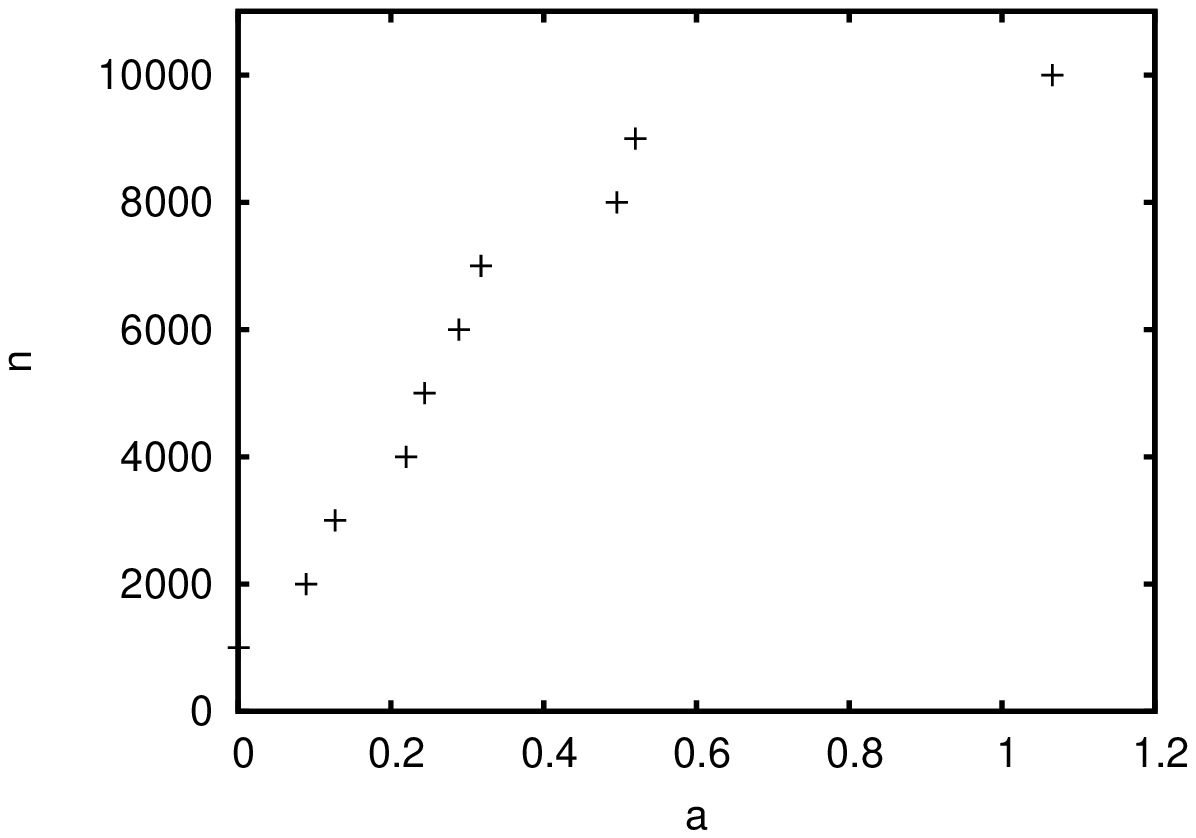}
%\vspace{0.4cm}
%\hspace{0.85cm}
\includegraphics[angle=0,width=3.6in]{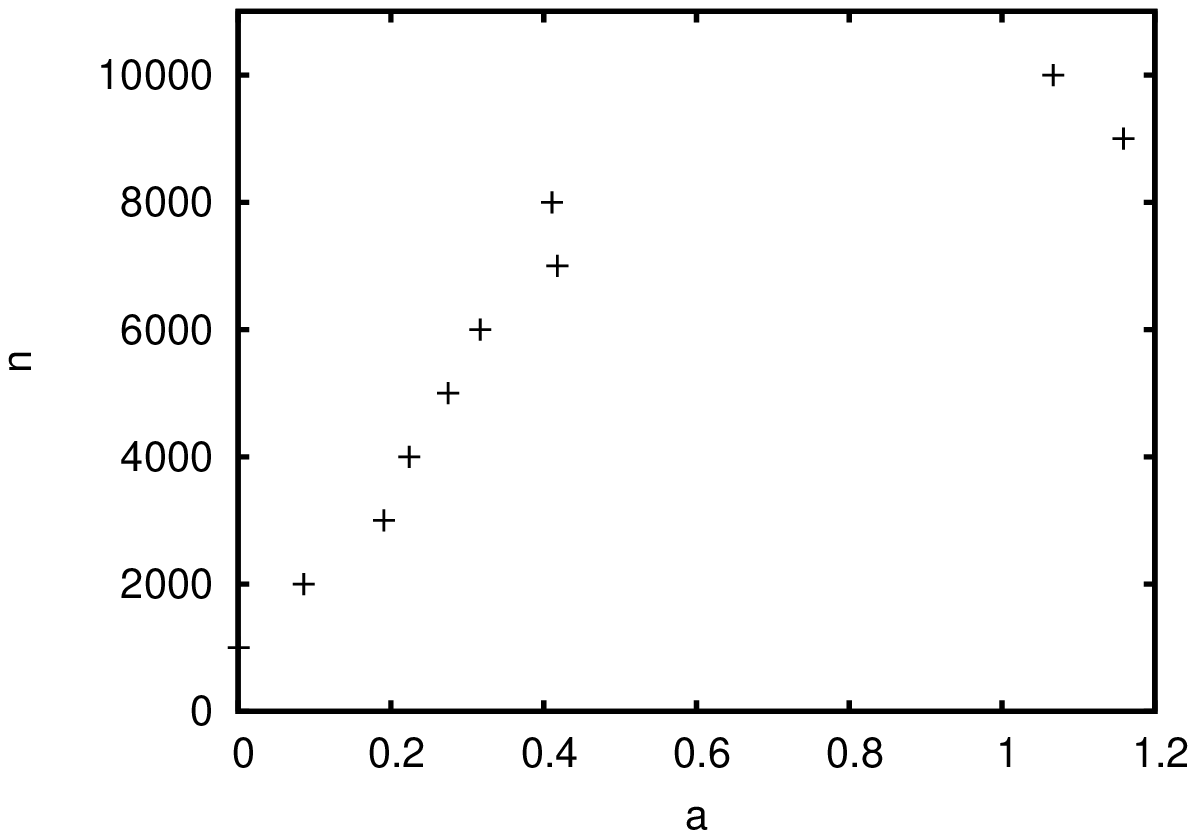}
%\vspace{0.4cm}
\caption{
\label{fig:mig}
The total distance ($x$-axis) migrated by the outer planet after 4000 orbits
as a function of number of planetesimals in the disk ($y$-axis). 
a) As measured from simulations C1-C10 which compute all pair-wise forces.
b) As measured from simulations N1-N10 which neglect the self-gravity
of the planetesimals.
For these simulations the planetesimals have the same mass so the particle
number sets the total disk mass.  We find that
the outermost planet migrates faster in proximity to a more
massive planetesimal disk at rate that
is approximately proportional to the disk density.   
The planetary migration rate is
on average $\sim 1.3$ times higher when disk self-gravity is neglected.
The lines show the best fit to the data points.
}
\end{figure}

We notice differences in the migration rates in the self-gravitating
disk compared to that lacking self-gravity.   On average the total
distance traveled by the outer planet is higher when self-gravity
is absent, as shown in Figure \ref{fig:mig} showing the total
distance traveled by the outer planet after 4000 orbits.  
However when we examine the total distance traveled
by the outer planet, we 
note that the self-gravitating simulations have much
larger scatter in this quantity than those lacking self gravity.
The planets in the simulations lacking self-gravity tend to have smoother
migration excepting when the disk is very massive.

For the most massive disks considered we note that the
migration rate strongly increases with time for both
self-gravitating and non-self gravitation simulations
(see Figure \ref{fig:mig_rate_profile}).
These may represent migration in the exponential or ``forced''
regime predicted under some conditions by \citet{gomes04}.

Following the toy model by \citet{gomes04}
(see their equations 1-3)
the planet migration rate, $\dot{a}_p$ is expected to depend on 
the total mass of orbit crossing planetesimals $M(t)$
\begin{equation}
\dot{a}_p \approx {k \over 2\pi} {M(t) \over M_p } {1 \over \sqrt{a_p}},
\end{equation}
where $k$ is a possibly migration rate dependent function 
that depends on the distribution of the planet crossing planetesimal orbits,
$M_p$ is the mass of the planet, and 
$a_p$ is the planet's semi-major axis. 
The evolution of $ M(t)$ depends on the timescale, $\tau$, for the planet to
scatter planetesimals away so that they are no longer orbit crossing
(also possibly migration rate dependent)
and the additional mass in planetesimals that becomes orbit crossing because
the planet has moved further into the unscattered disk;
\begin{equation}
\dot{M}(t) = - M(t)/\tau + 2 \pi a_p |\dot{a}_p| \sigma(a_p),
\label{eqn:mdot}
\end{equation}
where $\sigma(a)$ is the surface density of the unscattered 
planetesimal disk.

The sign of the parameter
\begin{equation}
\alpha = - \tau^{-1} + k \sqrt{a_p} \sigma(a_p)/M_p  
\end{equation}
determines whether the migration is damped or forced. 
If $\alpha <0$ then $M(t)$ decays exponentially and the
planet stops migrating, otherwise the planet accelerates.
For the most massive disks considered here we do see accelerations
in the migration rate, and for the lowest density
disks we see little migration.  However for most of our simulations we
see smooth or nearly powerlaw migration rates. 
For the migration rate to fail to be exponential
we require that $\alpha \approx 0$. There
could be an intermediate 
of disk densities, in between the forced and damped regimes,
where $k$ and $\tau$ can be considered
functions of the migration rate.   
Alternatively the migration could increase
until a saturation level which depends on disk density.
We note that the disks we consider are more massive
than required for migration; $\sigma(r)r^2 \ga M_p$.
For the massive disks a larger population of scattered
objects is left after the planet passes through the disk
implying that only a fraction of it has efficiently imparted
energy and angular momentum to the planet.
The planet scatters particles at lower efficiency
during more rapid migration.  This
would lower $k$ and $\tau$ possibly accounting
for the near cancellation of the two terms comprising $\alpha$.

%\subsection{Migration dependence on particle size}

\begin{figure}
%\vspace{-1.5cm}
%\hspace{.85cm}
\includegraphics[angle=0,width=3.6in]{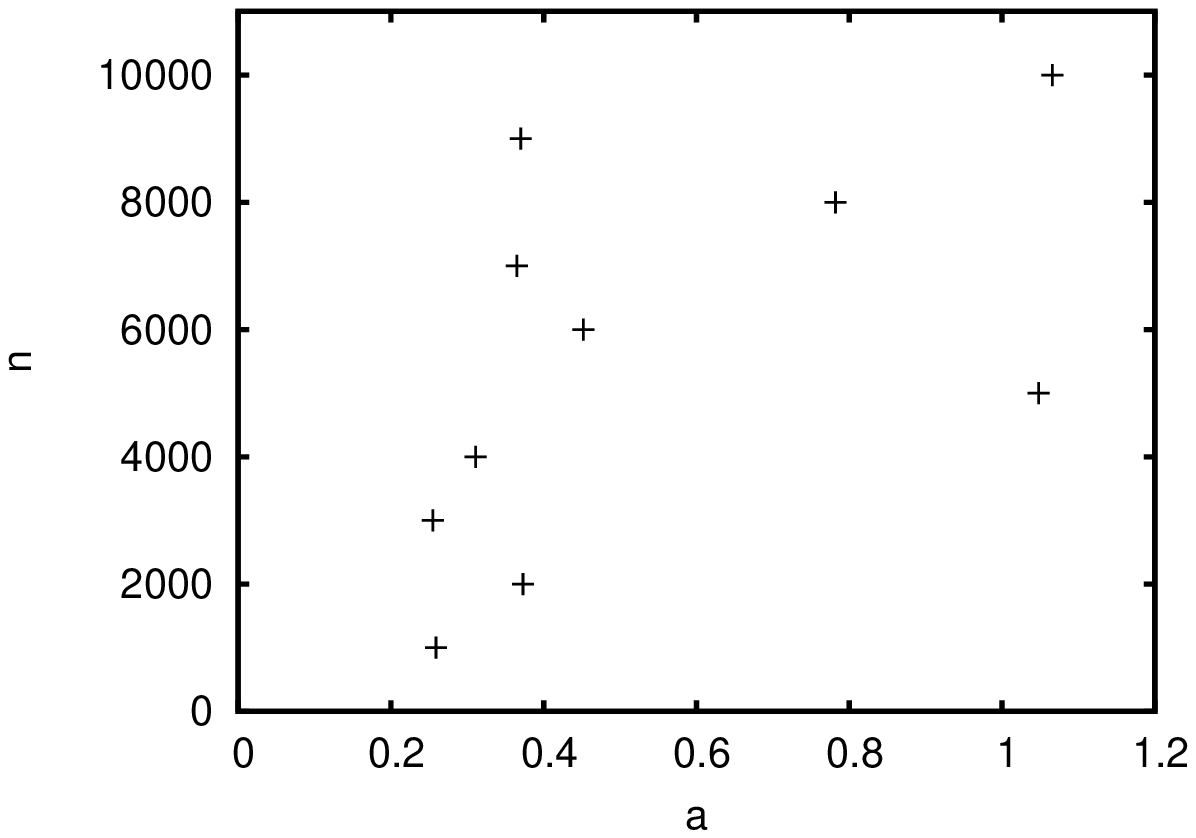}
%\vspace{0.4cm}
%\hspace{.85cm}
\includegraphics[angle=0,width=3.6in]{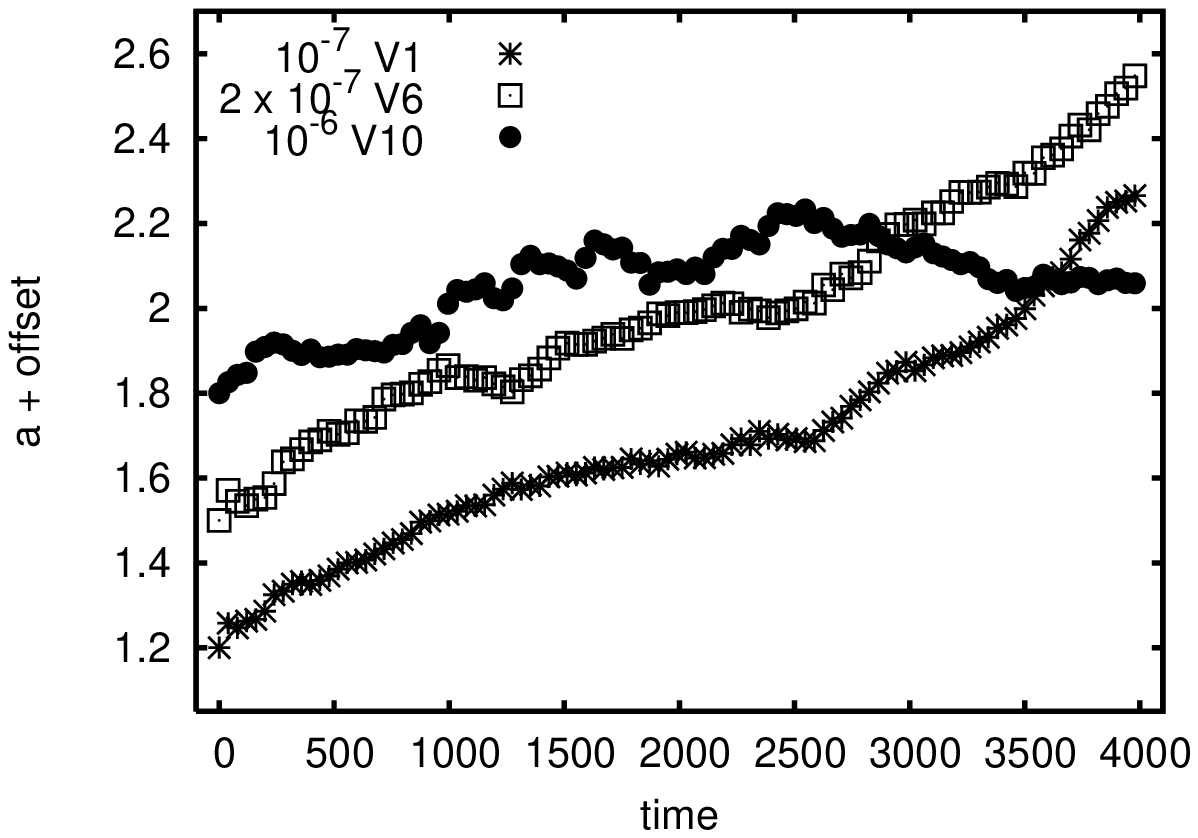} 
%\vspace{0.75cm}
\caption{
\label{fig:vmass}
a) The total distance traveled by the outer planet 
in 4000 orbits as a function
of the number of planetesimals in the disk for simulations V1-V10.
For these simulations the total disk mass is held fixed so that
the the planetesimal number is inversely proportional to their mass.
The total distance traveled by the outer planet is not strongly
dependent on the mass of the planetesimals. 
b) The semi-major axis as a function of time in orbits for
simulations V1, V6, and V10, with the same
disk mass but different masses and numbers of
planetesimals.   
The three sets of
data points have semi-major axes offset by 0.3 so the three
trajectories can be seen together on the same plot.
The key lists the simulations and planetesimal mass ratios for those
simulations.
The planet motion is more
stochastic when the planetesimals are more massive.
}
\end{figure}

We now fix the mass of the disk and consider how the mass of the particles
affect the migration rates.
In the V1-V10 simulations we altered the mass of the planetesimals such that
no matter what number of particles there where, the total disk mass 
was always a Jupiter mass.  
Figure \ref{fig:vmass} shows the distance traveled by the outer most planet 
and semi-major axis as a function of time for 3 simulations.
While Figure \ref{fig:vmass}a shows a substantial variation
in the total distanced traveled by the planets in these simulations
but no strong dependence on the planetesimal mass.
Figure \ref{fig:vmass}b, showing the semi-major axis
profiles shows that the migration of the outer planet is
far more stochastic when the planetesimal mass is higher
as explored by previous studies \citep{hahn99,zhou02,murray06}.

%Conclusion
\section{Summary and Discussion}

In this paper we have described an implementation
of the heliocentric democratic 2nd order symplectic 
integration scheme \citep{duncan98}
written in CUDA and designed to work on a graphics processing unit.
This allows us to efficiently use the large 
number of processors available on the card in parallel.
The code is designed 
to compute a large number of gravitational interactions as well
as compute Keplerian advances in parallel.
Because all 
memory resides on the device this parallel computation platform has advantages
over message passage interfaces that must pay a penalty to share information 
between processors.  The biggest drawback with our current computational 
platform is the low precision, a problem that is now being resolved with
the recent availability of double precision capable video cards.

We have used the new code to 
simulate planetary migration into a planetesimal 
disk with $10^4$ bodies and have computed all gravitational interactions.  
We have explored the difference between simulations of
planets migrating into self-gravitating planetesimal disks
and those lacking computed interactions between planetesimals.
We find that the fraction of particles
present in mean motion resonances is reduced in
the simulations with self-gravitating disks compared
to those lacking planetesimal interactions.  We attribute
this to a reduction in the resonance capture rate
due to the larger eccentricity dispersion caused
by gravitational stirring in the self-gravitating planetesimal disks.
This suggests that disks in proximity to migrating planets
could be featureless (as is true for Fomalhaut, \citealt{kalas05}).
A dust disk lacking resonant structure could still host
migrating planets.

We find that planet migration is smoother but somewhat faster in the simulations
lacking self-gravity.  Migration rates in both cases are approximately
linearly dependent on planetesimal disk density.
The planet can undergo rapid changes in migration in the simulations with
the most massive disks.  
However for the most of the disk densities considered 
no exponential increase in migration rate is seen
implying that the migration rate saturates at a level
approximately proportional to the disk density.

\vskip 1.0 truein
%\acknowledgments
We thank Matt Holman for suggesting how to reduce the radial
drift caused by single precision computation during the Keplerian
computation step.
Support for this work at University of Rochester and Rochester Institute of Technology
was also provided by NASA through an award issued by JPL/Caltech, 
by NSF grants AST-0406823 \& PHY-0552695 and 
and HST-AR-10972 to the Space Telescope Science Institute.
This work is based on observations made with the Spitzer Space Telescope, which is operated
by the Jet Propulsion Laboratory, California Institute of Technology under a contract with NASA.

{}

\begin{table*}
\begin{minipage}{120mm}
\caption{Simulations }
\label{tab:tab1}
\begin{tabular}{@{}llcccccc}
\hline
Simulation & $M_1$& $a_1$& $M_2$     	     &$a_2$& $m$       &  $N$           & $M_d$ \\ 
\hline
C1     &$10^{-3}$ & 1.0  & $5\times 10^{-5}$ & 1.5 & $10^{-7}$ & $10^4$         & $10^{-3}$ \\
C2     &$10^{-3}$ & 1.0  & $5\times 10^{-5}$ & 1.5 & $10^{-7}$ & $9\times 10^3$ & $9\times 10^{-4}$\\
C3     &$10^{-3}$ & 1.0  & $5\times 10^{-5}$ & 1.5 & $10^{-7}$ & $8\times 10^3$ & $8\times 10^{-4}$\\
C4     &$10^{-3}$ & 1.0  & $5\times 10^{-5}$ & 1.5 & $10^{-7}$ & $7\times 10^3$ & $7\times 10^{-4}$\\
C5     &$10^{-3}$ & 1.0  & $5\times 10^{-5}$ & 1.5 & $10^{-7}$ & $6\times 10^3$ & $6\times 10^{-4}$\\
C6     &$10^{-3}$ & 1.0  & $5\times 10^{-5}$ & 1.5 & $10^{-7}$ & $5\times 10^3$ & $5\times 10^{-4}$\\
C7     &$10^{-3}$ & 1.0  & $5\times 10^{-5}$ & 1.5 & $10^{-7}$ & $4\times 10^3$ & $4\times 10^{-4}$\\
C8     &$10^{-3}$ & 1.0  & $5\times 10^{-5}$ & 1.5 & $10^{-7}$ & $3\times 10^3$ & $3\times 10^{-4}$\\
C9     &$10^{-3}$ & 1.0  & $5\times 10^{-5}$ & 1.5 & $10^{-7}$ & $2\times 10^3$ & $2\times 10^{-4}$\\
C10    &$10^{-3}$ & 1.0  & $5\times 10^{-5}$ & 1.5 & $10^{-7}$ & $10^3$ 		& $10^{-4}$\\
N1     &$10^{-3}$ & 1.0  & $5\times 10^{-5}$ & 1.5 & $10^{-7}$ & $10^4$         & $10^{-3}$ \\
N2     &$10^{-3}$ & 1.0  & $5\times 10^{-5}$ & 1.5 & $10^{-7}$ & $9\times 10^3$ & $9\times 10^{-4}$\\
N3     &$10^{-3}$ & 1.0  & $5\times 10^{-5}$ & 1.5 & $10^{-7}$ & $8\times 10^3$ & $8\times 10^{-4}$\\
N4     &$10^{-3}$ & 1.0  & $5\times 10^{-5}$ & 1.5 & $10^{-7}$ & $7\times 10^3$ & $7\times 10^{-4}$\\
N5     &$10^{-3}$ & 1.0  & $5\times 10^{-5}$ & 1.5 & $10^{-7}$ & $6\times 10^3$ & $6\times 10^{-4}$\\
N6     &$10^{-3}$ & 1.0  & $5\times 10^{-5}$ & 1.5 & $10^{-7}$ & $5\times 10^3$ & $5\times 10^{-4}$\\
N7     &$10^{-3}$ & 1.0  & $5\times 10^{-5}$ & 1.5 & $10^{-7}$ & $4\times 10^3$ & $4\times 10^{-4}$\\
N8     &$10^{-3}$ & 1.0  & $5\times 10^{-5}$ & 1.5 & $10^{-7}$ & $3\times 10^3$ & $3\times 10^{-4}$\\
N9     &$10^{-3}$ & 1.0  & $5\times 10^{-5}$ & 1.5 & $10^{-7}$ & $2\times 10^3$ & $2\times 10^{-4}$\\
N10    &$10^{-3}$ & 1.0  & $5\times 10^{-5}$ & 1.5 & $10^{-7}$ & $10^3$ 		& $10^{-4}$\\
V1     &$10^{-3}$ & 1.0  & $5\times 10^{-5}$ & 1.5 & $10^{-7}$ 				& $10^4$         & $10^{-3}$ \\
V2     &$10^{-3}$ & 1.0  & $5\times 10^{-5}$ & 1.5 & $1.085\times 10^{-7}$ 	& $9\times 10^3$ & $10^{-3}$ \\
V3     &$10^{-3}$ & 1.0  & $5\times 10^{-5}$ & 1.5 & $1.25\times 10^{-7}$ 	& $8\times 10^3$ & $10^{-3}$ \\
V4     &$10^{-3}$ & 1.0  & $5\times 10^{-5}$ & 1.5 & $1.4\times 10^{-7}$ 	& $7\times 10^3$ & $10^{-3}$ \\
V5     &$10^{-3}$ & 1.0  & $5\times 10^{-5}$ & 1.5 & $1.67\times 10^{-7}$ 	& $6\times 10^3$ & $10^{-3}$ \\
V6     &$10^{-3}$ & 1.0  & $5\times 10^{-5}$ & 1.5 & $2.0\times 10^{-7}$ 	& $5\times 10^3$ & $10^{-3}$ \\
V7     &$10^{-3}$ & 1.0  & $5\times 10^{-5}$ & 1.5 & $2.5\times 10^{-7}$ 	& $4\times 10^3$ & $10^{-3}$ \\
V8     &$10^{-3}$ & 1.0  & $5\times 10^{-5}$ & 1.5 & $3.25\times 10^{-7}$ 	& $3\times 10^3$ & $10^{-3}$ \\
V9     &$10^{-3}$ & 1.0  & $5\times 10^{-5}$ & 1.5 & $5\times 10^{-7}$ 		& $2\times 10^3$ & $10^{-3}$ \\
V10    &$10^{-3}$ & 1.0  & $5\times 10^{-5}$ & 1.5 & $10^{-6}$ 				& $10^3$ 		 & $10^{-3}$ \\
%E1     &$10^{-3}$ & 1.0  & $5\times 10^{-5}$ & 2.0 & $10^{-7}$ & $10^4$ 		& $10^{-4}$\\
%E2     &$10^{-3}$ & 1.0  & $1\times 10^{-4}$ & 2.0 & $10^{-7}$ & $10^4$ 		& $10^{-4}$\\
\hline
\end{tabular}
{ \\
Here $M_1$, $M_2$ are the masses of the two planets, 
and $m$ is the planetesimal mass in units of the central star mass. 
The parameters $a_1$ and $a_2$ are the initial planet semi-major axes.
The parameter $N$ is the total number of planetesimals, and $M_d$ is
the total planetesimal disk mass in units of the stellar mass.
The planetesimals initially extend between 1.5 and 3.0 in semi-major axis.
Simulations C1-C10 included all pair-wise forces all have the same
planetesimal masses but a range of total disk masses.
Simulations N1-N10 are identical 
to the C series but neglect interactions between the planetesimals.
Simulations V1-V10 compute all pair-wise forces and vary planetesimal mass in order to maintain constant mass disk with a fixed number of particles.
%Simulations E1 and E2 were simulations run with planetesimals embedded at a semi major axis of 2.0.
Simulations were run for approximately 4000 orbits.
}
\end{minipage}
\end{table*}


\begin{thebibliography}{}

%Hybrid N-body-coagulation code for planet formation
\bibitem[Bromley \& Kenyon(2006)]{bromley06}
Bromley, B. C., \& Kenyon, S. J. 2006, AJ, 131, 2737

%Age Dependence of the Vega Phenomenon: Theory
\bibitem[Dominik \& Decin(2003)]{DD03}
Dominik, C., \& Decin, G. 2003, ApJ, 598, 626

% A multiple timestep symplectic algorithms for integrating close encounters
\bibitem[Duncan et al.(1998)]{duncan98}
Duncan, M. J.,  Levison, H. F., \& Lee, M. H. 1998, AJ, 116, 2067

% Simplectic Correctors
\bibitem[Wisdom, Holman \& Touma(1996)]{wisdom96}
Wisdom, J.,  Holman, M., \& Touma, J. 1996, 
Fields Institute Communications, 10, 217

%   Some dynamical aspects of the accretion of Uranus and Neptune 
% - The exchange of orbital angular momentum with planetesimals
\bibitem[Fernandez \& Ip(1984)]{fernandez84}
Fernandez, J. A., \& Ip, W.-H. 1984, Icarus, 58, 109

%Planetary migration in a planetesimal disk: why did Neptune stop at 30 AU?
\bibitem[Gomes et al.(2004)]{gomes04}
Gomes, R. S., Morbidelli, A., \& Levison, H. F.	
2004, Icarus, 170, 492

%Origin of the cataclysmic Late Heavy Bombardment period of the terrestrial planets
\bibitem[Gomes et al.(2005)]{gomes05}
Gomes, R., Levison, H. F., Tsiganis, K., \& Morbidelli, A. 2005, 
Nature, 435, 466

%Neptune's Migration into a Stirred-Up Kuiper Belt: A Detailed Comparison of Simulations to Observations
\bibitem[Hahn \& Malhotra(2005)]{hahn05}
Hahn, J. M., Malhotra, R.  2005, AJ, 130, 2392


%Orbital Evolution of Planets Embedded in a Planetesimal Disk
\bibitem[Hahn \& Malhotra(1999)]{hahn99}
Hahn, J. M., \& Malhotra, R. 1999, AJ, 117, 3041

%Parallel Prefix Sum (scan) with CUDA
\bibitem[Harris et al.(2008)]{harris08}
Harris,  M., Sengupta, S, \& Owens, J. D.   2008, Chap 39 
in GPUGems3, edited by Hubert Nguyen,
2008, Addison-Wesley,  Upper Saddle River, NJ, page 851

%A planetary system as the origin of structure in Fomalhaut's dust belt
\bibitem[Kalas et al.(2005)]{kalas05}
Kalas, P., Graham, J. R., \& Clampin, M. 2005, Nature,  435, 1067

%Orbital Migration of Neptune and orbital distribution of trans-Neptunian objects
\bibitem[Ida et al.(2000)]{ida00}
Ida, S., Bryden, G., Lin, D. N. C., \& Tanaka,  H. 2000, ApJ, 534, 428

%\bibitem[Juric \& Tremaine(2007)]{juric07} 
%Juric, M., \& Tremaine, S. 2007, ApJ, submitted (astro-ph /0703160)

%Orbital Evolution of Protoplanets Embedded in a Swarm of Planetesimals
\bibitem[Kokubo \& Ida(1995)]{kokubo95}
Kokubo, E., \& Ida, S. 1995, Icarus, 114, 247

%The formation of the Kuiper belt by the outward transport of bodies during Neptune's migration
\bibitem[Levison \& Morbidelli(2003)]{levison03}
Levison, H. F., Morbidelli, A. 2003, Nature, 426, 419


%Origin of the structure of the Kuiper belt during a dynamical instability in the orbits of Uranus and Neptune
\bibitem[Levison et al.(2008)]{levison08}
Levison, H. F., Morbidelli, A., Vanlaerhoven, C., Gomes, R., \& Tsiganis, K.\
2008, Icarus, 196, 258	

% forced migration, no interactions at all
%planetsimals in the precense of giant planet migration
% however point out signature of planet migration in remaining scattered population
\bibitem[Lufkin et al.(2006)]{lufkin06}
Lufkin, G., Richardson, D. C., \& Mundy, L. G. 2006,
ApJ, 653, 1464

%The Origin of Pluto's Orbit: Implications for the Solar System Beyond Neptune
\bibitem[Malhotra(1995)]{malhotra95}
Malhotra, R. 1995, AJ, 110, 420

%Modern integrations of Solar System dynamics
\bibitem[Morbidelli(2001)]{morby01}
Morbidelli, A. 2001,
Ann. Rev. Earth and Pl. Sci., 30, 89 %-112                             

%Brownian motion in Planetary migration
\bibitem[Murray-Clay \& Chiang(2006)]{murray06}
Murray-Clay, R. A., \& Chiang, E. I. 
2006, ApJ, 651, 1194

% Fast N-body simulation with CUDA
\bibitem[Nyland, Harris, \& Prins(2008)]{nyland08}
Nyland, L.,  Harris,  M., \& and Prins, J.  2008, Chap 31 
in GPUGems3, edited by Hubert Nguyen,
2008, Addison-Wesley,  Upper Saddle River, NJ, page 677

\bibitem[Prussing \& Conway(1993)]{prussing93}
Prussing, J. E. \& Conway, B. A.  1993, Orbital Mechanics, Oxford
University Press, Inc., New York, New York

%Planetary embryos and planetesimals residing in thin debris discs
\bibitem[Quillen et al.(2007)]{quillen07}
Quillen, A. C., Morbidelli, A., \& Moore, A. 2007, MNRAS, 380, 1642

%Predictions for a planet just inside Fomalhaut's eccentric ring
\bibitem[Quillen(2006a)]{quillen06}
Quillen, A. C.  2006, MNRAS, 372, L14

%Reducing the probability of capture into resonance
\bibitem[Quillen(2006b)]{quillen06b}
Quillen, A. C. 2006, MNRAS, 365, 1367

%\bibitem[Rasio \& Ford(1996)]{rasio96}
%Rasio, F. A., \& Ford, E. B. 1996, Science, 274, 954

% from mmrs to scattered planets
\bibitem[Thommes et al.(2008)]{thommes08}
Thommes, E. W., Bryden, G., Wu, Y., \& Rasio, F. A. 2008,
ApJ, 675, 1538

\bibitem[Tsiganis et al.(2005)]{tsiganis05}
Tsiganis, K., Gomes, R., Morbidelli, A., \& Levison, H. F. 2005, Nature, 435, 459

%Collisional processes in extrasolar planetesimal discs - dust clumps in Fomalhaut's debris disc
\bibitem[Wyatt \& Dent(2002)]{wyatt02}
Wyatt, M. C., \& Dent, W. R. F. 2002, MNRAS, 334, 589

%Dust in Resonant Extrasolar Kuiper Belts: Grain Size and Wavelength Dependence of Disk Structure
\bibitem[Wyatt(2006)]{wyatt06}
Wyatt, M. C 2006, ApJ, 639, 1153

\bibitem[Yoshida(1990)]{yoshida90}
Yoshida, H. 1990, Phys. Lett. A, 150, 262

%Stochastic effects in the planet migration and orbital distribution
%of the Kuiper Belt
\bibitem[Zhou et al.(2002)]{zhou02}
Zhou, L.-Y., Sun, Y.-S. Zhou, J.-L., Zheng, J.Q., \& Valtonen, M.
2002, MNRAS, 336, 520


\end{thebibliography}
\end{document}